\documentclass[9pt, conference]{IEEEtran}
\IEEEoverridecommandlockouts

\usepackage{cite}
\usepackage{bm}

\usepackage{amsmath,amssymb,amsfonts}
\usepackage{algorithmic}
\usepackage{graphicx}
\usepackage{hyperref}
\usepackage{textcomp}
\usepackage{adjustbox}
\usepackage{xcolor}
\usepackage{booktabs}
\usepackage{authblk}

\def\BibTeX{{\rm B\kern-.05em{\sc i\kern-.025em b}\kern-.08em
    T\kern-.1667em\lower.7ex\hbox{E}\kern-.125emX}}

\usepackage{circledsteps}
\newcommand\myCircled[2][]{\ifmmode%
\Circled[fill color=black,inner color=white,#1]{\footnotesize\mathsf{#2}}%
\else%
\Circled[fill color=black,inner color=white,#1]{\footnotesize\sffamily#2}%
\fi%
}

\newcommand{\PacketCLIP}{\textsc{PacketCLIP} }

\begin{document}

\title{\textsc{PacketCLIP}: Multi-Modal Embedding of Network Traffic and Language for Cybersecurity Reasoning}

\author[1]{Ryozo Masukawa}
\author[1]{Sanggeon Yun}
\author[1]{Sungheon Jeong}
\author[1]{Wenjun Huang}
\author[1]{Yang Ni}
\author[1]{Ian Bryant} 
\author[2]{\\Nathaniel D. Bastian}
\author[1]{Mohsen Imani\thanks{Corresponding author, email: m.imani@uci.edu}}

\affil[1]{University of California, Irvine}
\affil[2]{Army Cyber Institute, United States Military Academy}

\maketitle

\begin{abstract}
Traffic classification is vital for cybersecurity, yet encrypted traffic poses significant challenges. We present \PacketCLIP, a multi-modal framework combining packet data with natural language semantics through contrastive pretraining and hierarchical Graph Neural Network (GNN) reasoning. \PacketCLIP integrates semantic reasoning with efficient classification, enabling robust detection of anomalies in encrypted network flows. By aligning textual descriptions with packet behaviors, it offers enhanced interpretability, scalability, and practical applicability across diverse security scenarios. \PacketCLIP achieves a 95\% mean AUC, outperforms baselines by 11.6\%, and reduces model size by 92\%, making it ideal for real-time anomaly detection. By bridging advanced machine learning techniques and practical cybersecurity needs, \PacketCLIP provides a foundation for scalable, efficient, and interpretable solutions to tackle encrypted traffic classification and network intrusion detection challenges in resource-constrained environments.

\end{abstract}


\vspace{-4mm}
\section{Introduction}

\begin{figure*}[t!]
    \centering
    \includegraphics[width=\linewidth]{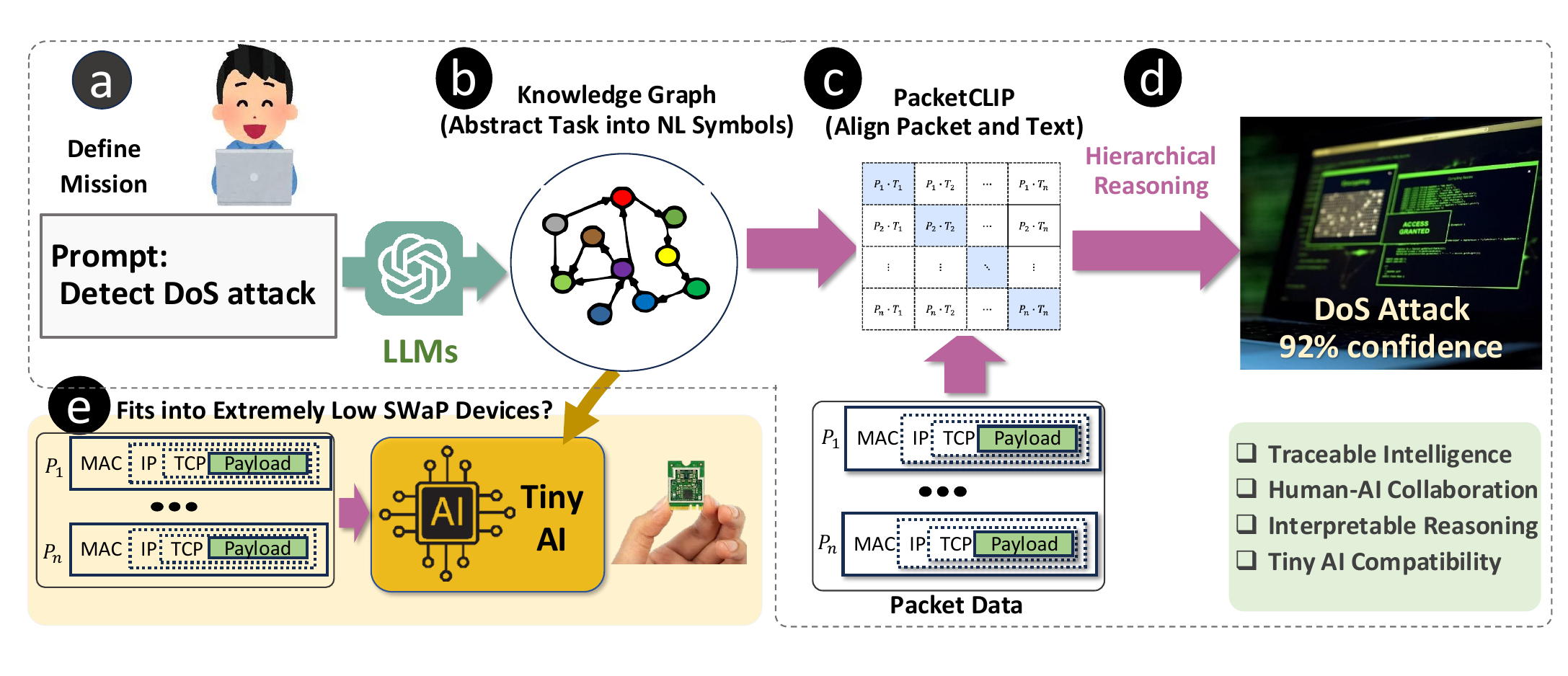}
    \vspace{-11mm}
    \caption{Semantic AI framework for detecting traffic related to specific cyber-attacks defined by a user(\myCircled{a}), combining LLM-driven knowledge graphs(\myCircled{b}), \PacketCLIP alignment(\myCircled{c}), hierarchical reasoning(\myCircled{d}), and Tiny AI to enable efficient, interpretable, and traceable detection on low-resource devices(\myCircled{e}).}
    \label{fig:overview}
    \vspace{-5mm}
\end{figure*}

\vspace{-2mm}
Traffic classification plays a crucial role in modern network security analytics, significantly influencing areas such as threat detection and micro-segmentation strategies. As networks become increasingly dynamic, the ability to accurately classify traffic is essential for enhancing security and swiftly responding to business needs. Traditionally, traffic classification techniques relied on inspecting packet headers and payloads. However, the rise of encrypted and anonymized traffic presents significant challenges by obscuring content, making it harder to distinguish between benign and malicious flows.

Recent advances in machine learning, particularly with pre-trained models based on architectures like BERT~\cite{devlin2018bert, lin_et-bert_2022, dac04_nlp_hardware_security} and masked autoencoders (MAEs)~\cite{zhao2023yet}, have attempted to address this issue and achieved state-of-the-art performance in various security-related tasks, including encrypted traffic classification. Hyperdimensional Computing (HDC)-based hardware-efficient methods have also been proposed~\cite{dac_hdc}. These methods leverage deep learning to identify patterns in packet metadata and encrypted content, bypassing the need for payload inspection. Despite their technical advancements, these models still struggle to track and interpret the semantics of network behaviors, particularly when trying to discern the underlying intent or strategy of cyberattacks. Reasoning about the semantics of cyberattacks remains a key research challenge.

In the field of video anomaly detection, MissionGNN~\cite{yun2024missiongnn}, a cutting-edge hierarchical graph neural network (GNN) model, has demonstrated exceptional capability in reasoning about anomalies using mission-specific knowledge graphs (KGs). By incorporating node embeddings derived from dual modalities—natural language and image data—and employing joint-embedding models such as CLIP~\cite{pmlr-v139-radford21a}, MissionGNN effectively reasons across both visual and textual domains. This success prompts an intriguing question: \textit{Can this hierarchical GNN-based reasoning be adapted for encrypted traffic detection, and if so, could it address the persistent challenge of semantic reasoning in the cybersecurity domain?} This question arises from the conceptual similarity between video (a sequence of images) and network flows (a sequence of packets), which, while differing in modality, share a sequential structure.

To explore this possibility, we propose a semantic reasoning framework for encrypted traffic detection, illustrated in \autoref{fig:overview}. In our framework, the user first defines a specific task in encrypted traffic detection (e.g., detecting a Denial of Service (DoS) attack as shown in \autoref{fig:overview}~\myCircled{a}). Then, a Large Language Model (LLM) generates a KG, which is a Directed Acyclic Graph (DAG) serving as an abstract representation of the task to be used later as a medium for reasoning (\autoref{fig:overview}, \myCircled{b}).

To adapt hierarchical GNN reasoning to encrypted traffic detection, we need a joint-embedding model capable of mapping both encrypted traffic data and Natural Language (NL) into a unified vector space (\autoref{fig:overview}, \myCircled{c}). This approach allows us to leverage the semantic reasoning capabilities of hierarchical GNNs while addressing the unique challenges posed by encrypted network traffic (\autoref{fig:overview}, \myCircled{d}). To enable alignment between text and packet data, we propose \textbf{\PacketCLIP}, which utilizes recent advancements in LLMs~\cite{openai2024gpt4technicalreport, touvron_llama_2023} to create a multi-modal joint embedding via contrastive pretraining. Inspired by Contrastive Language–Image Pre-training (CLIP), which links images with text, \PacketCLIP connects packet-level traffic data with semantic descriptions. This alignment not only improves traffic classification accuracy but also provides human operators with NL explanations of packet behavior within the network flow, enhancing interpretability.

We conducted experiments to evaluate the effectiveness and efficiency of \PacketCLIP in conjunction with hierarchical GNN reasoning. The results demonstrate that \PacketCLIP achieves not only high classification accuracy but also significant improvements in robustness and efficiency. Specifically, hierarchical reasoning with \PacketCLIP delivers an impressive $\mathbf{11.6\%}$ mean Area Under the Curve (mAUC) improvement compared to baseline methods. Notably, it maintains $\mathbf{95\%}$ mAUC performance when trained on just $30\%$ of the data, significantly outperforming ET-BERT, which achieves only $50\%$ mAUC under the same conditions. In addition to its performance advantages, \PacketCLIP is highly efficient, achieving a $\mathbf{92\%}$ reduction in the number of parameters and a $\mathbf{98\%}$ reduction in FLOPs compared to existing methods. These efficiency gains underscore the model's ability to deliver strong performance with a significantly smaller computational footprint.  Overall, these findings highlight \PacketCLIP's capability to generalize effectively in data-constrained scenarios and its suitability for practical deployment in environments with limited computational resources.

Finally, we evaluated the performance of both \PacketCLIP and its hierarchical GNN within a real-time traffic intrusion detection framework using the ACI-IoT-2023 dataset~\cite{qacj3x3223}. Our results show that \PacketCLIP effectively aligns packet and text modalities, while the hierarchical GNN achieves robust and energy-efficient intrusion detection. Because our GNN-based reasoning framework is intended for practical deployment in routers (\autoref{fig:overview}, \myCircled{e}), these results underscore its potential for real-world applications. The key contributions of this research are as follows:
\begin{itemize}
    \item Proposed \PacketCLIP, a multi-modal framework aligning encrypted traffic data with NL.
    \item Introduced contrastive pretraining and hierarchical GNN reasoning for robust intrusion detection, outperforming baselines by $\mathbf{11.6\%}$ in AUC scores.
    \item Showed strong data scarcity resilience, maintaining $\mathbf{95\%}$ mAUC even with $\mathbf{30\%}$ training data, compared to $\mathbf{70\%}$ for ET-BERT.
    \item Achieved $\mathbf{92\%}$ parameter and $\mathbf{98\%}$ FLOPs reduction, enabling deployment in resource-constrained environments.
    \item Validated on real-world datasets, combining robust traffic classification with efficient and scalable anomaly detection for practical network security applications.
\end{itemize}

\section{Background and Related Works}

This section aims to highlight advancements in traffic intrusion detection, outlining significant progress while identifying ongoing challenges. By examining modern approaches, particularly those leveraging machine learning and GNNs, we underscore the field’s evolution and remaining challenges in achieving effective, privacy-preserving, and interpretable detection techniques. We also emphasize the differences between our proposed method and previous works.

\begin{figure*}[t!]
    \centering    \includegraphics[width=1.0\linewidth]{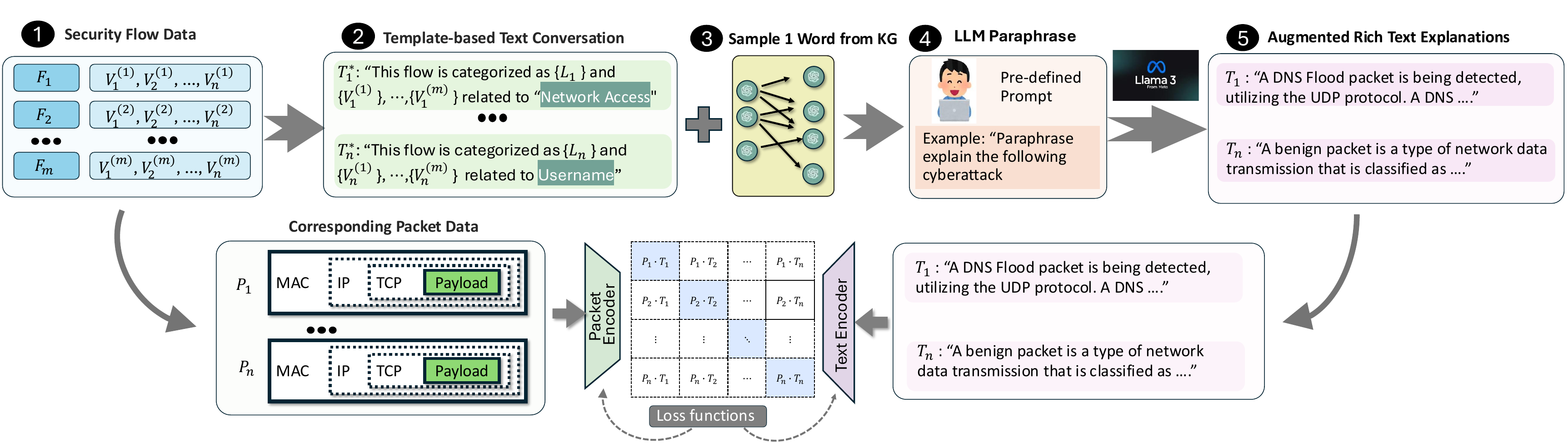}
    \vspace{-4mm}
    \caption{A framework to generate NL explanations for intrusion scenarios by mapping tabular security flow data(\myCircled{1}) to text templates (\myCircled{2}), leveraging LLM-generated knowledge graphs (\myCircled{3}), utilizing LLMs for paraphrased explanations (\myCircled{4}), and producing interpretable descriptions of network events (\myCircled{5}).}
    \label{fig:nl_expression}
    \vspace{-5mm}
\end{figure*}

Port-based classification methods~\cite{port_based}, which historically provided effective means for categorizing network traffic, have encountered limitations due to dynamic port allocations, rendering it difficult to track application-specific patterns accurately. Traditional deep packet inspection (DPI) techniques~\cite{papadogiannaki2021survey}, which analyze data payloads for distinguishing patterns, have similarly become impractical, especially for encrypted traffic. The computational burden and diminishing returns on accuracy for DPI methods, as encryption becomes more widespread, highlight the critical need for machine learning-driven approaches that accommodate complex traffic patterns while maintaining privacy.
Statistical feature-based approaches leverage manually selected traffic features, requiring substantial domain expertise~\cite{197185, panchenko2016website, zaki2022grain, 7467370}. For instance, AppScanner~\cite{7467370} utilizes statistical attributes of packet sizes to train random forest classifiers; however, these methods suffer from limitations in capturing high-level semantic patterns essential for robust intrusion detection. Recent works have introduced GNN-based frameworks for enhanced traffic classification, leveraging graph structures to capture relational dependencies within traffic data~\cite{huoh_flow-based_2023, zhang2023tfe, zhangone2024, dac03_gnn_security}. Among them, TFE-GNN~\cite{zhang2023tfe} is notable for modeling packet payloads at the byte level, treating each byte as a node and creating edges based on point-wise mutual information (PMI) between nodes. Additionally, a novel contrastive learning-based intrusion detection framework extending TFE-GNN has shown promising results. However, these GNN-based methods often struggle with interpretability, as they rely on encrypted byte representations that do not lend themselves to human understanding. Consequently, these methods may fall short in supporting human security analysts in devising precise micro-segmentation policies.

In the field of video anomaly detection, MissionGNN~\cite{yun2024missiongnn} has demonstrated state-of-the-art performance by employing KG reasoning techniques and shows powerful following works. 
Building upon this approach, we introduce a novel framework that combines a GNN-based reasoning component with \PacketCLIP, a cross-modal embedding model designed to align packet data with NL descriptions within a shared vector space. To the best of our knowledge, this is the first integration of NL processing with GNN-based network traffic intrusion detection, facilitating intuitive and interpretable reasoning over encrypted traffic patterns. 
\vspace{-3mm}
\section{Methodology}

\subsection{Mission-specific Knowledge Graph Generation}

First, to enable traffic classification and reason about the semantics of the attack, the mission-specific KG generation framework is used to create a KG that extracts relevant information from a given packet. In encrypted traffic detection, each mission-specific KG represents structured knowledge about a particular event or scenario. The process begins by obtaining a set of vocabularies for each event, referred to as \textbf{Key Concepts}. This is done using an LLM such as GPT-4o~\cite{openai2024gpt4technicalreport} with the prompt: \textit{"List} $V(\in \mathbb{N})$ \textit{typical vocabularies to represent [event name]? Note: Everything should be a single word."} This generates a first key concept set, denoted as $K$.

Next, we expand $K$ by querying the LLM with the prompt: \textit{"What are associated words with vocabularies in set $K$?"} This produces a set of associated vocabularies $K^{(i)}_{a}$ for each key concept, ensuring no overlap with the original set ($K \cap K^{(i)}_{a} = \emptyset$). The set $K$ is updated using the equation:

\begin{equation}
    K = K \cup K^{(i)}_{a} \quad (1 \leq i \leq N)
\end{equation}

This process is repeated for $N$ iterations, after which edges are drawn from the $(i-1)$th key concept to the $i$th key concept, forming a hierarchical directed acyclic graph (DAG).

On top of the mission-specific KG, a sensor node is added, containing sensory information such as packet data encoded by joint-embedding models like \PacketCLIP. Directed edges are projected from the sensor node to key concept nodes, and related concept nodes also project edges to an embedding node, which aggregates messages passed through the graph.

This KG design allows the GNN to pass interpretable messages by embedding multimodal information from all nodes into a unified vector space thanks to the \PacketCLIP alignment.
\vspace{-4mm}
\subsection{NL Explanation for Intrusion}

A key challenge in our framework is achieving a rich textual representation of cyberattacks, as, to the best of our knowledge, no existing datasets related to network traffic classification include NL descriptions. Most current datasets (~\cite{qacj3x3223, neto2023ciciot2023, draper2016characterization, dadkhah2022towards}) generally provide two main types of data: (1) raw packets stored in PCAP files with corresponding labels and (2) tabular data representing network flows derived from these PCAP files, typically in CSV format (see \myCircled{1} in \autoref{fig:nl_expression}).
Our approach focuses specifically on the tabular flow files, as certain columns within this data have the potential to serve as elements in generating NL descriptions for each packet’s semantic context. As shown in \autoref{fig:nl_expression}, we first convert the tabular data into a template-based text expression (Step \myCircled{2}) by embedding each column value into structured sentences.

To train \PacketCLIP with a rich and diverse vocabulary comparable to LLMs, we incorporate a mission-specific KG that aligns with each tabular data row's label, such as those associated with DoS attack detection. We enhance textual variety by sampling nodes from the KG and integrating them into template-based descriptions (Step \myCircled{3}). Recognizing the limitations of static templates, we employ lightweight, open-source LLMs to paraphrase these descriptions, ensuring grammatical correctness and semantic variety for each cyberattack (Step \myCircled{4}). Finally, we can obtain a diverce text expression of each packet (Step \myCircled{5}). This augmentation not only mitigates the constraint of limited class labels but also elevates the diversity of template-based learning, enabling broader generalization.

This NL augmentation method we described so far can be formulated as follows. We first obtain $n (\in \mathbb{N})$ template-based text data ($\mathcal{D}^{*}_{T} = \{\mathbf{T}^{*}_{i}\}_{i=1}^{n}$) from flow that is represented as tabular data. After this, we use LLMs as follows 
\begin{equation}
    \mathbf{T}_{i} = LLM(\mathbf{T}^{*}_{i}) (1 \leq i \leq n),
\end{equation}
where $LLM$ indicates LLMs and we obtain augmented highly diverse text expression of each packet.

\subsection{\PacketCLIP Contrastive Pretraining}
\begin{figure*}[t!]
    \centering
    \includegraphics[width=\linewidth]{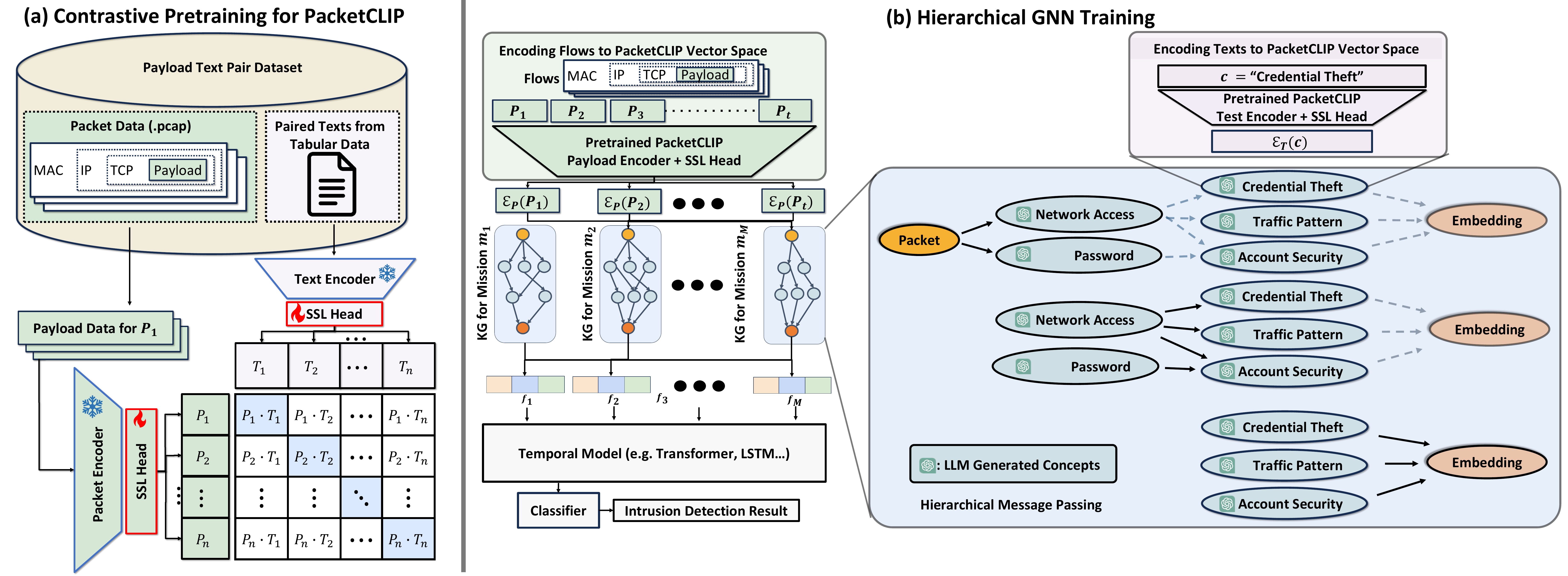}
    \vspace{-4mm}
    \caption{(a) The overall architecture of the contrastive pretraining process for \PacketCLIP, including encoding packets and paired texts for learning.\\ (b) A mission-specific hierarchical GNN framework that integrates \PacketCLIP with temporal models and classifiers to derive intrusion detection results.}
    \vspace{-3mm}
    \label{fig:pretrainig}
\end{figure*}

Using the approach described in the previous section, \PacketCLIP obtains a set of text data, $\mathcal{D}_{T} = \{\mathbf{T}_{i}\}_{i=1}^{n}$, paired with corresponding packet payload data, $\mathcal{D}_{P} = \{\mathbf{p}_{i}\}_{i=1}^{n}$. We clarify the pretraining of \PacketCLIP described in \autoref{fig:pretrainig} (a).

During contrastive pretraining, we keep the weights of both the text encoder ($f_{T}$) and the packet encoder ($f_{P}$) fixed. This strategy preserves the consistency of text representations and avoids catastrophic forgetting~\cite{french1999catastrophic}. Rather than fine-tuning the large pretrained models used for the packet and text encoders, we adopt a method from ~\cite{gupta2022understanding}, introducing a simple linear projection layer as a self-supervised learning (SSL) head for each encoder: one for the packet encoder ($g_{P}$) and another for the text encoder ($g_{T}$). During pretraining, only these projection layers are updated.

The text $\mathbf{t} \in \mathcal{D}_{T}$ and its paired packet $\mathbf{p} \in \mathcal{D}_{P}$ are encoded as follows:
\begin{equation}
\mathbf{z_{t}} = g_{T} \circ f_{T}(\mathbf{t}), \quad \mathbf{z_{p}} = g_{P} \circ f_{P}(\mathbf{p}).
\end{equation}

For contrastive pretraining, we use the InfoNCE loss~\cite{pmlr-v119-chen20j} $l$ as defined below:
\begin{equation}
l(\mathbf{z}_{t}, \mathbf{z}_{p}; \mathcal{Z}^{\backslash}) = - \log{\frac{\exp\left(\cos(\mathbf{z_{t}}, \mathbf{z_{p}^{+}})/\tau\right)}{\sum_{\mathbf{z_{p}}^{\backslash} \in \mathcal{Z}^{\backslash}} \exp\left(\cos(\mathbf{z_{t}}, \mathbf{z_{p}^{\backslash}})/\tau\right)}},
\end{equation}
where $\mathcal{Z}^{\backslash}$ denotes a set of embedded vectors sampled from $\mathcal{D}_{P}$ that excludes the packet vector ($\mathbf{z}_{p}^{+}$) matching the text vector $\mathbf{z_{t}}$, and $\tau > 0$ represents the temperature parameter. This loss function encourages alignment between embeddings from paired text and packet instances while pushing apart embeddings from different instances.

By completing this contrastive pretraining process, \PacketCLIP learns robust, aligned representations for both text and packet data, enhancing its ability to capture semantic connections between textual and packet-based cyberattack data.

\subsection{Hierarchical GNN Reasoning}


After generating $M$ KGs $G_{m_i}$ ($1 \leq i \leq M$, where $m_i$ denotes the $i$-th mission.), we train a hierarchical GNN model to classify events or anomalies in network traffic data (\autoref{fig:pretrainig} (b)).
GNNs capture relational information using feature vectors for each node, connecting packet node features \( x^{(0)}_{s, m_i} \) for packet at timestamp $t$ ($P_{t}$) from the packet encoder \( \mathcal{E}_P (= g_{P} \circ f_{P}) \) and concept node features \( x^{(0)}_{c, m_i} \) for each concept $c$ from the text encoder \( \mathcal{E}_T (=g_{T} \circ f_{T}) \) as follows:

\begin{equation}
    \bm{x}^{(0)}_{s, m_i} = \mathcal{E}_P(P_t), \hspace{3mm} \bm{x}^{(0)}_{c, m_i} = \mathcal{E}_T(c)
\end{equation}

A multi-layer perceptron (MLP) then embeds these node features at layer $l (1\leq l \leq L)$ of GNN as follows:

\begin{equation}
    \bm{x}^{(l)}_{m_i} = W^{(l)}_{m_i} \bm{x}^{(l-1)}_{m_i} + \bm{b}^{(l)}_{m_i}
\end{equation}

where $W^{(l)}_{m_i}$ denotes a trainable weight matrix and $\bm{b}^{(l)}_{m_i}$ indicates the bias. The core idea is hierarchical message passing, where messages are propagated through three levels of the KG hierarchy: packet nodes to key concepts, key concepts to associated concept nodes, and finally to embedding nodes. This structure allows efficient and targeted aggregation of information across modalities, resulting in interpretable, goal-oriented embeddings.

Hierarchical message passing from node $v$ to neighboring node in the previous hierarchy $u$ at layer $l$ is recursively defined as:
\vspace{-1mm}
\begin{equation}
    \bm{x}^{(l)}_v = \frac{1}{|\mathcal{N}^{(h-1)}(v)|} \sum_{u \in \mathcal{N}^{(h-1)}(v)} \phi^{(l)}(\bm{x}^{(l-1)}_v \cdot \bm{x}^{(l-1)}_u)
\end{equation}
\vspace{-1mm}
where $\phi$ indicates the activation functon and $\mathcal{N}^{(h)}(v)$ represents the neighbors of node $v$ at hierarchy $h$. The final embeddings feature node $\bm{x}_{emb}$ for each mission-specific KG are combined into a single vector:

\begin{equation}
    \bm{f}^{(t)} = [\bm{x}^{(L)}_{\text{emb}, m_1}, \bm{x}^{(L)}_{\text{emb}, m_2}, \ldots, \bm{x}^{(L)}_{\text{emb}, m_M}]
\end{equation}

For each packet $F_t$, the sequence of tokens \( X_t \) is constructed as:

\vspace{-3mm}
\begin{equation*}
    X_t = \{\bm{f}^{(t-A+1)}, \bm{f}^{(t-A+2)}, \ldots, \bm{f}^{(t)}\}
\end{equation*}

where $A$ represents a hyperparameter that specifies a fixed number of time frames to be input into the temporal model. This sequence is input into a Transformer encoder $\mathcal{T}$ followed by an MLP to produce the final classification output:

\begin{equation}
    \hat{\bm{y}} = \text{Softmax}(\text{MLP}(\mathcal{T}(X_t)))
\end{equation}

Training leverages cross-entropy loss, smoothing loss, and anomaly localization techniques to optimize the GNN model for network traffic event recognition and anomaly detection tasks following ~\cite{yun2024missiongnn}.

\section{Experiments}

\begin{figure*}[ht]
    \centering
    \begin{minipage}{0.36\textwidth}
        \includegraphics[width=\linewidth]{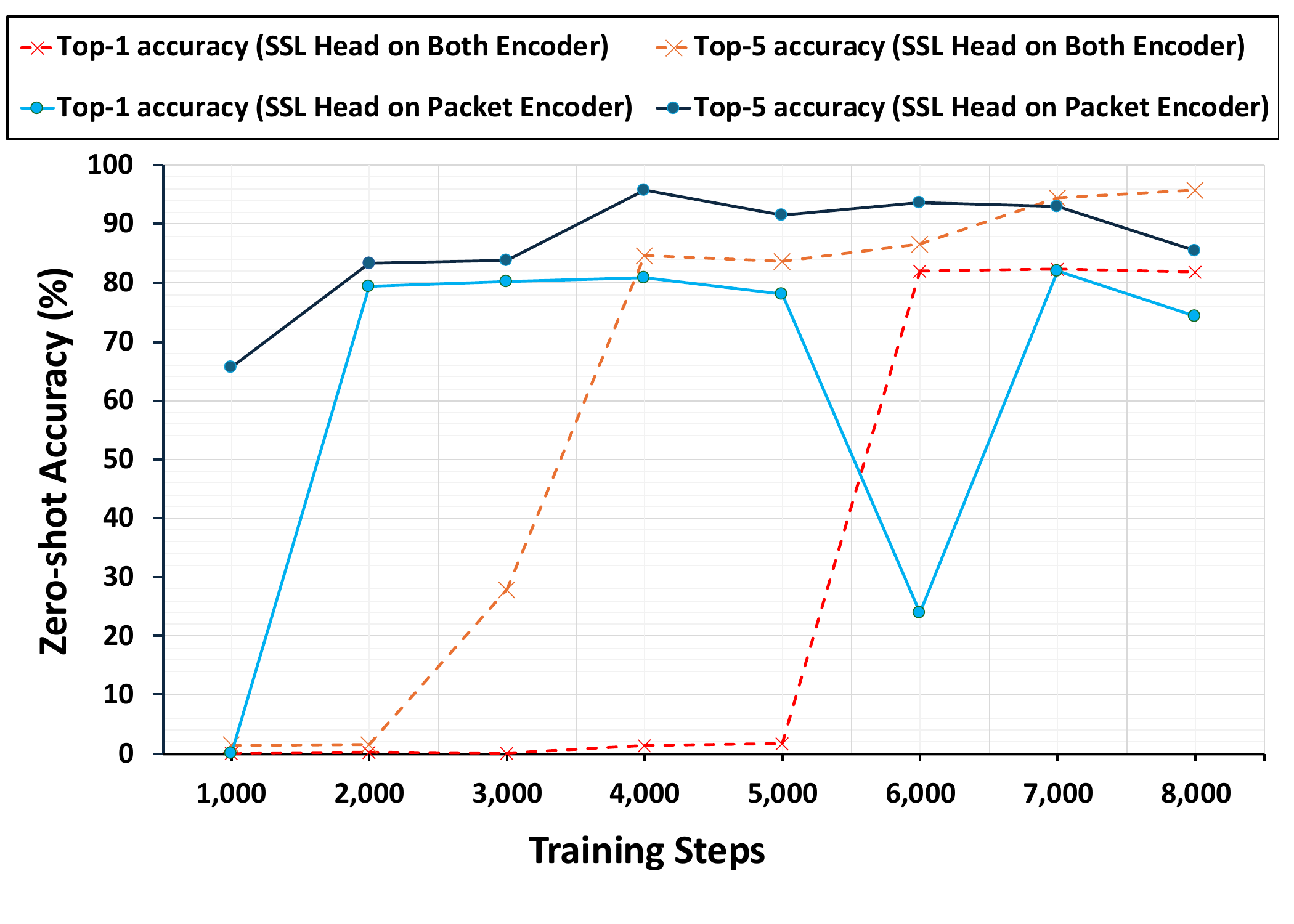}
    \caption{Zero-shot accuracy change during training shows a trade-off: SSL on both encoders improves faster but is less stable, while SSL only on the packet encoder progresses slower but is more stable.}
    \label{fig:zero-shot}
    \end{minipage}
    \hfill
    \begin{minipage}{0.6\textwidth}
        \includegraphics[width=\linewidth, height=80mm,keepaspectratio]{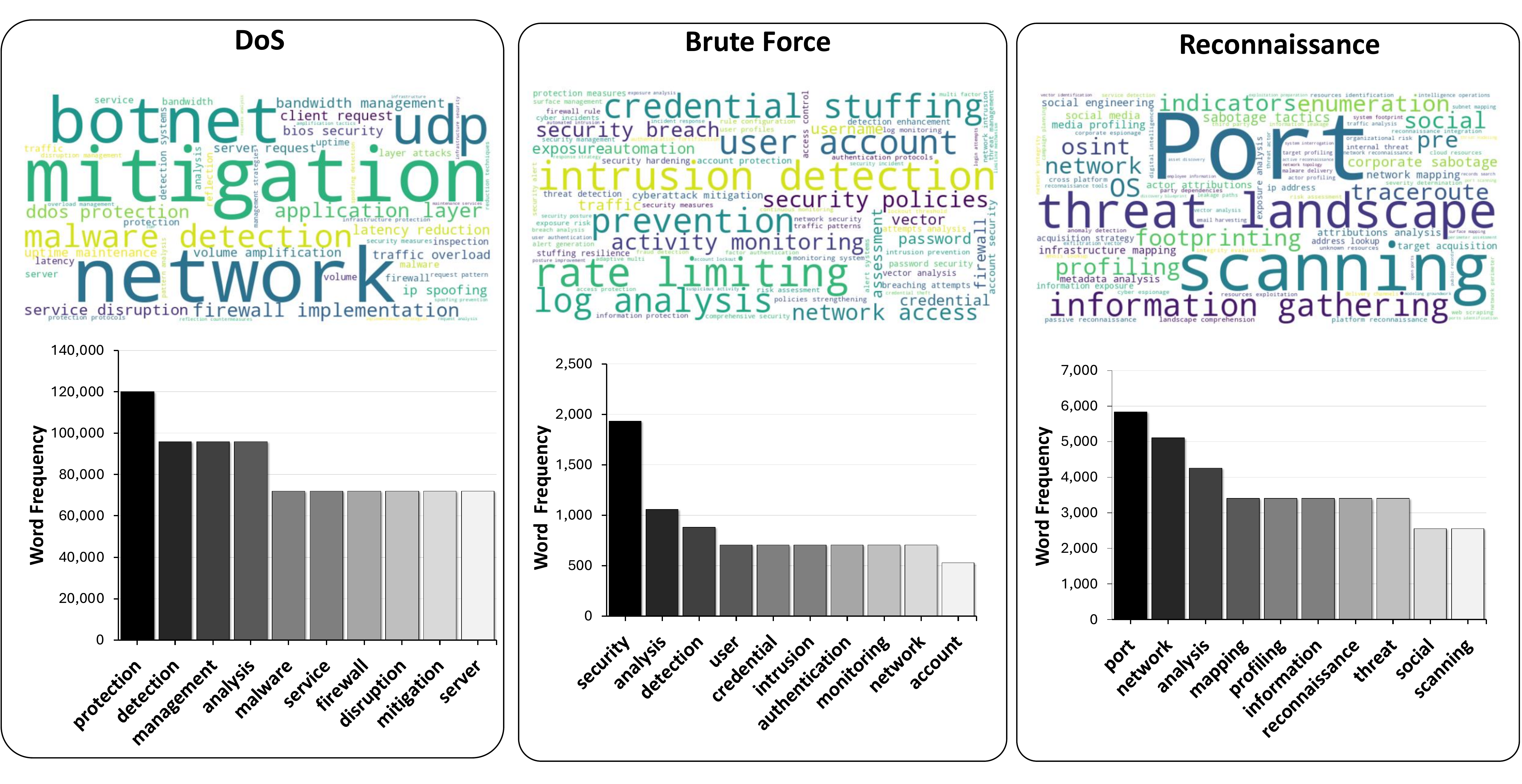}
        \vspace{-7mm}
        \caption{Word clouds and top 10 frequent vocabularies for DoS, Brute Force, and Reconnaissance missions from ACI-IoT-2023, highlighting key terms like 'botnet,' 'credential,' and 'port scanning' for respective categories.}
        \label{fig:aci-iot-text}
    \end{minipage}
    \label{fig:both_figures}
    \vspace{-3mm}
\end{figure*}

\begin{table}[t!]
    \centering
    \caption{Comparison of models highlighting \PacketCLIP configurations achieving highest top-1 and top-5 performance metrics.}
    \label{tab:cmp_desc}
    \begin{adjustbox}{width=0.48\textwidth}
    \begin{tabular}{c|c|c}
    \toprule
    Method& top-1 & top-5\\
    \midrule
    \textit{ET-BERT}~\cite{lin_et-bert_2022} & 0.730 & - \\
    \PacketCLIP (No SSL Head) & 0.001 & 0.955  \\
    \PacketCLIP (SSL Head Only on Packet Encoder) & 0.831 & \textbf{0.991} \\
    \PacketCLIP (SSL Head on both Encoder) & \textbf{0.856} & 0.961 \\
    \bottomrule
    \end{tabular}%
    \end{adjustbox}
\end{table}
\begin{table*}[ht]
\centering
\caption{AUC scores for individual classes and their mean AUC (mAUC) are compared across models. The results highlight the performance superiority of \PacketCLIP with GNN Reasoning, achieving the highest scores in all categories.}
\vspace{-3mm}
\label{tab:auc_scores}
\begin{adjustbox}{width=0.8\textwidth}
\begin{tabular}{l|cccc|c}
\toprule
\textbf{Model} & \textbf{Benign} & \textbf{DoS} & \textbf{Reconnaissance} & \textbf{Brute Force} & \textbf{Average} \\
\midrule

\textit{ET-BERT} Fine-Tuning\cite{lin_et-bert_2022}  & 0.717 & 0.731 & 0.752 & 0.784 & 0.746 \\
\PacketCLIP  + XGBoost & 0.522 & 0.500 & 0.477 & 0.458 & 0.489 \\
\PacketCLIP + LightGBM & 0.961 & 0.544 & 0.974 & 0.567 & 0.761 \\
\PacketCLIP + DNN &  0.978 & 0.724 & 0.996 & 0.623 & 0.830 \\
\PacketCLIP + GNN Reasoning & \textbf{0.996} & \textbf{0.930}& \textbf{0.999}&  \textbf{0.909} & \textbf{0.946} \\
\bottomrule
\end{tabular}
\end{adjustbox}
\vspace{-5mm}
\end{table*}

\subsection{Implementation details}
For converting tabular data into NL expressions, we used LLaMA 3~\cite{touvron_llama_2023}. For KG generation, we employed an automated framework powered by GPT-4o~\cite{openai2024gpt4technicalreport}. The \PacketCLIP packet encoder was implemented using the \textit{ET-BERT} pretrained encoder~\cite{lin_et-bert_2022}, while the text encoder relied on RoBERTa~\cite{liu2019roberta}. For optimization, we adopted the Adam~\cite{diederik2014adam} optimizer, configured with a learning rate of \( 5.0 \times 10^{-4} \), \(\beta_{1} = 0.9\), \(\beta_{2} = 0.8\), and \(\epsilon = 1.0 \times 10^{-6}\). Within the hierarchical GNN model, we ensured a consistent dimensionality of \( D_{m_i, l} = 8 \) for mission \( m_i \) at hierarchy \( l \). For the short-term temporal model, an internal dimensionality of 128 was employed, alongside 8 attention heads and the hyperparameter $A$ was set to $30$. The training process was conducted over 3,000 steps, utilizing a mini-batch size of 128 samples for each step.

\vspace{-2mm}
\subsection{Datasets}
We utilized the ACI-IoT-2023 dataset~\cite{qacj3x3223}, a comprehensive IoT cybersecurity dataset containing labeled benign and malicious traffic, including threats like malware, DoS, and botnets. Using an LLM-based paraphrasing method (\autoref{fig:nl_expression}), we generated diverse payload-text pairs to enhance semantic representation.
For \PacketCLIP pretraining evaluated in \autoref{sec:top1}, data was classified into 10 distinct categories: \textit{Benign}, \textit{OS Scan}, \textit{Vulnerability Scan}, \textit{Port Scan}, \textit{ICMP Flood}, \textit{Slowloris}, \textit{SYN Flood}, \textit{UDP Flood}, \textit{DNS Flood}, and \textit{Dictionary Attack}. 

For GNN-based reasoning classification evaluated in \autoref{sec:gnn}, we grouped the dataset into broader categories: \textit{Benign}, \textit{DoS}, \textit{Reconnaissance}, and \textit{Brute Force}.  We treated the data as time-series packet sequences by attaching timestamps and sorting chronologically, allowing GNN models to leverage temporal structures similar to video data analysis.

\subsection{Visualization of LLM generated cyberattack semantics}
\autoref{fig:aci-iot-text} shows the textual explanations generated from the ACI-IoT dataset, emphasizing the most frequent terms used to describe network events. The lower bar graph highlights common words and phrases, such as “attack,” “network,” and “security,” which encapsulate key cybersecurity themes. Above, word clouds visually represent mission-specific vocabularies, showing the terms that form the nodes of corresponding KGs. Together, these visualizations illustrate \PacketCLIP’s ability to generate contextually relevant explanations, providing enriched semantic insights into various cyber incidents.
\vspace{-2mm}
\subsection{\PacketCLIP Semantic Classification Performance}
\label{sec:top1}
\noindent \textbf{Baselines}: To establish a baseline, we fine-tuned the \textit{ET-BERT}~\cite{lin_et-bert_2022} packet classifier on the ACI-IoT-2023 dataset, demonstrating the effectiveness of leveraging NL-based semantics for improved classification performance. Additionally, we performed an ablation study to evaluate the contribution of the SSL head in \PacketCLIP. Specifically, we examined three configurations: \PacketCLIP without any SSL head, with a single SSL head applied only to the packet encoder, and with SSL heads applied to both the text and packet encoders. We did not include an analysis of applying a single SSL head to the text encoder because the primary goal of \PacketCLIP's contrastive learning is to align the non-interpretable packet modality with the NL modality. Applying an SSL head solely to the text encoder could potentially degrade valuable NL information, which is counterintuitive to our objective.

\noindent \textbf{Evaluation metrics} : For evaluation on zero-shot performance, we used both top-1 and top-5 accuracy. For comparison, since the baseline methods can only output a single classification result, we mainly compare with top-1 accuracy.

\autoref{tab:cmp_desc} shows the performance comparison between each baselines that \PacketCLIP methods perform better than \textit{ET-BERT} fine-tuning in traffic classification. While \textit{ET-BERT} demonstrates moderate accuracy, \PacketCLIP using only packet information achieves a noticeable improvement, with nearly perfect reliability when considering multiple predictions. \PacketCLIP, when incorporating both packet data and contextual information, further enhances its ability to make accurate top predictions but slightly reduces its broader predictive range. Overall, \PacketCLIP offers superior accuracy, especially when combining packet and contextual details, making it a more effective method for precise traffic classification in network management.
At the same time, \PacketCLIP's performance comparison in \autoref{fig:zero-shot}  highlights the critical role of the SSL head. When incorporated, the SSL head significantly boosts both top-1 and top-5 accuracy, demonstrating its ability to enhance classification reliability and precision. Without only having one SSL head in packet Encoder, the performance drops notably in the middle of the training, underlining its importance in leveraging SSL heads on both encoders effectively. This comparison underscores the value of the SSL head in extracting meaningful features from both packet data and contextual information, enabling more accurate predictions in traffic classification. The results clearly establish the SSL head as a crucial component for achieving superior classification performance in \PacketCLIP.

\begin{figure}[ht]
    \centering
    \includegraphics[width=\linewidth]{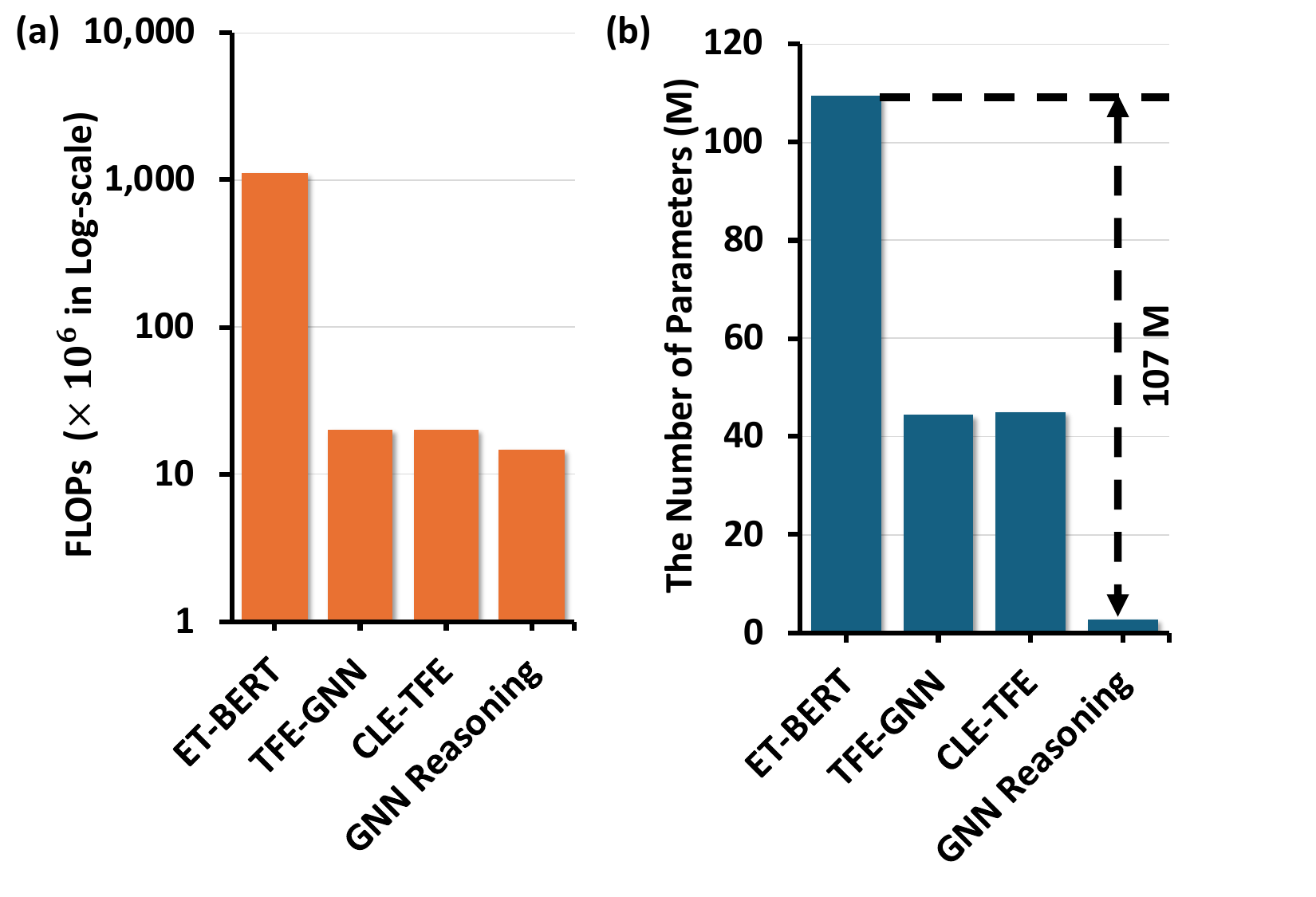}
    \vspace{-8mm}
    \caption{(a) Comparison of models based on FLOPs ($\times 10^{6}$ Log-scale) and (b) on the number of parameters (M).}
    \vspace{-2mm}
    \label{fig:hardware}
    \vspace{-4mm}
\end{figure}

\subsection{Evaluation on Hierarchical GNN Reasoning}
\label{sec:gnn}
\noindent \textbf{Baselines}: The ACI-IoT-2023 dataset, utilized in our experiments, has been previously explored in works such as AIS-NIDS~\cite{nate_baseline}. AIS-NIDS introduced a novel approach involving serialized RGB image transformations for packet-level feature extraction and employed basic machine learning models, including XGBoost and LightGBM, for intrusion detection. However, AIS-NIDS relies on closed-set classifiers and lacks publicly available code for its CNN preprocessing pipeline, posing challenges for reproducibility and adaptation to alternative methods. To address these limitations, we adopt baselines that incorporate the \PacketCLIP packet feature encoder in conjunction with various machine learning models. Specifically, we evaluate the following configurations: \PacketCLIP + XGBoost, \PacketCLIP + LightGBM, and \PacketCLIP (packet) + Deep Neural Network (DNN). Furthermore, to assess the performance of an external baseline, we fine-tuned \textit{ET-BERT}~\cite{lin_et-bert_2022} on the same dataset for comparative analysis.

\noindent \textbf{Evaluation metrics} : To evaluate our method, we adopted the AUC metric, which offers a robust measure of performance across all classification thresholds. AUC is particularly suitable for cybersecurity tasks, given the highly imbalanced nature of datasets, where attack instances are far fewer than benign traffic. By focusing on ranking instances correctly, AUC ensures a comprehensive evaluation of anomaly detection performance under varying conditions.

We present the result of our approach in \autoref{tab:auc_scores}, showing an average AUC score gain of more than $\mathbf{11.6\%}$ compared to baseline methods. This substantial improvement highlights significant advancements in intrusion detection, combining semantic reasoning, interpretability, and advanced classification capabilities. By employing \PacketCLIP alongside hierarchical GNN reasoning, we provide a robust and innovative solution tailored for real-time anomaly detection in IoT networks, demonstrating the potential for enhanced semantic understanding and improved adaptability compared to traditional methods.
\subsection{Hardware Efficiency Analysis}
The computational efficiency of the hierarchical GNN reasoning module in \PacketCLIP is compared against \textit{ET-BERT}~\cite{lin_et-bert_2022}, TFE-GNN~\cite{zhang2023tfe}, and CLE-TFE~\cite{zhangone2024} in terms of FLOPs and parameter count. These baselines are chosen due to their proven effectiveness in encrypted traffic detection tasks. 
\PacketCLIP’s GNN reasoning module demonstrates a significant improvement, achieving a 107M reduction in parameters compared to \textit{ET-BERT}, as shown in \autoref{fig:hardware} (b). This reduction highlights the module's streamlined architecture, which effectively aggregates semantic information through hierarchical message passing while minimizing parameterization. 
Moreover, as depicted in \autoref{fig:hardware} (a), the FLOPs of the GNN reasoning module are approximately one-thirtieth of \textit{ET-BERT}, while still delivering competitive performance in traffic anomaly detection tasks. 
These results emphasize the scalability of \PacketCLIP, particularly for resource-constrained environments like IoT networks, where computational overhead is a critical concern. By incorporating GNN reasoning, \PacketCLIP balances performance and efficiency, making it highly suitable for real-time intrusion detection applications.
\subsection{GNN Reasoning Robustness of Scarce Data Availability}

\begin{figure}
    \centering
    \includegraphics[width=\linewidth]{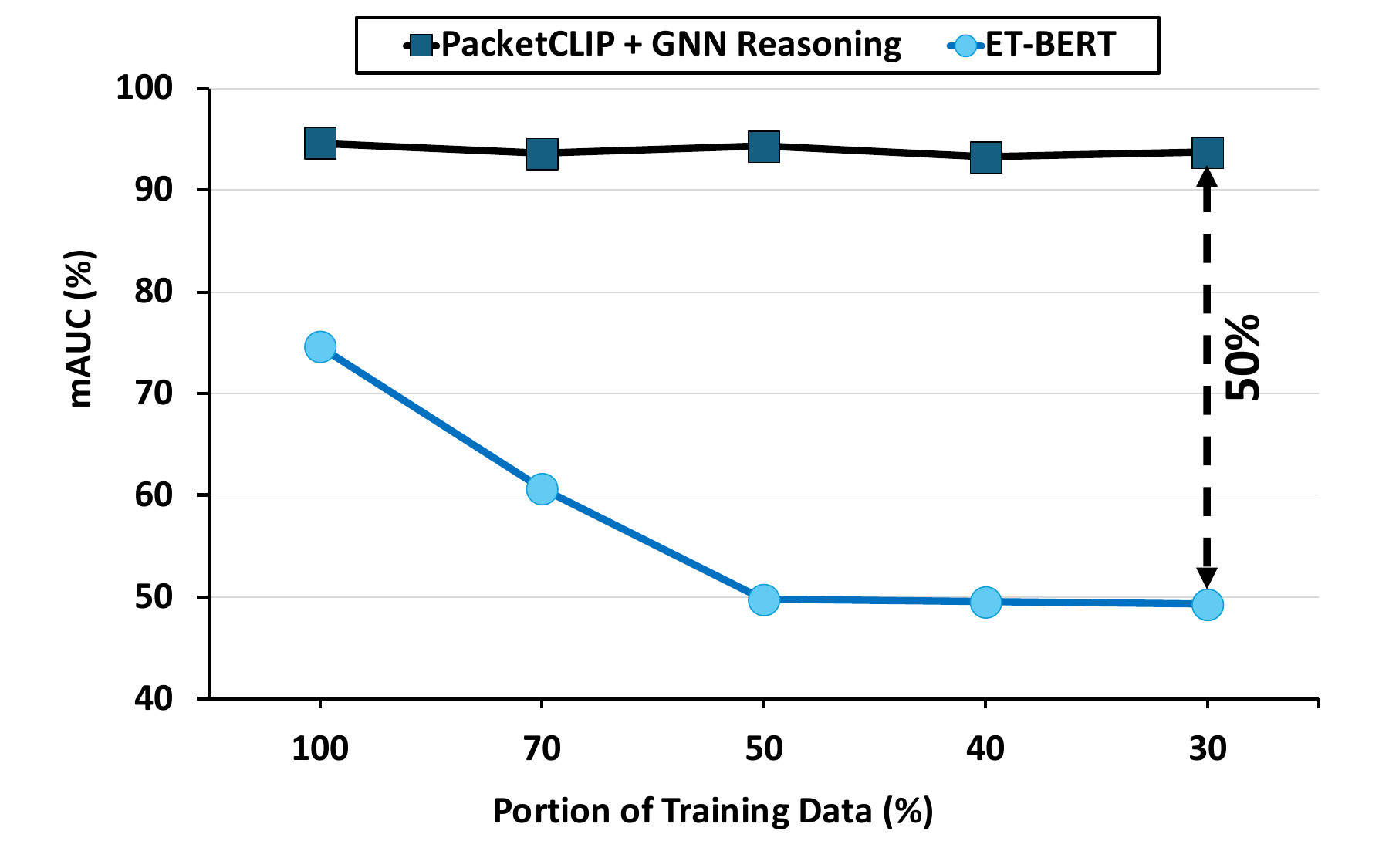}
    \vspace{-6mm}
    \caption{Robustness against Data Scarcity Analysis by mAUC Comparison of PacketCLIP + GNN Reasoning and \textit{ET-BERT} with Varying Training Data ($100\%, 70\%, 50\%, 40\%, 30\%)$. }
    \label{fig:sample-cmp}
    \vspace{-6mm}
\end{figure}

To assess the robustness of \PacketCLIP + GNN Reasoning under varying levels of data availability, we conducted experiments using the ACI-IoT-2023 dataset, selecting \textit{ET-BERT}~\cite{lin_et-bert_2022} as a baseline for comparison. Both models were trained on three different proportions of the training data: $100\%$, $70\%$, $50\%$, $40\%$, and $30\%$, while the test set remained consistent across all experiments to ensure a fair and controlled evaluation. The mAUC, again, served as the primary performance metric.

\autoref{fig:sample-cmp} presents the results of this experiment. \PacketCLIP + GNN Reasoning consistently achieved high mAUC scores ($\sim95\%$) across all training data splits, demonstrating strong generalization even with limited data. In contrast, \textit{ET-BERT} exhibited notable performance degradation, with mAUC dropping from $\sim 70\%$ at $100\%$ training data to $\sim50\%$ at $30\%$. These findings emphasize the robustness of \PacketCLIP + GNN Reasoning, making it well-suited for scenarios with constrained training data.

\vspace{-2mm}
\section{Conclusions}

We introduced \PacketCLIP, a multi-modal framework integrating packet-level data with NL semantics to advance encrypted traffic classification and intrusion detection. By combining contrastive pretraining and hierarchical GNN reasoning, \PacketCLIP demonstrates robustness in both performance and efficiency. It achieved an $11.6\%$ higher AUC compared to baseline models, consistently delivering superior mAUC scores of approximately $95\%$, even with only $30\%$ of the training data. These results highlight the resilience of hierarchical GNN reasoning in \PacketCLIP to handle data scarcity and its ability to generalize effectively.
Furthermore, \PacketCLIP reduces model size by $92\%$ and computational requirements by $98\%$, making it highly efficient for real-time applications in resource-constrained environments like IoT networks. By providing interpretable semantic insights alongside robust anomaly detection, \PacketCLIP harmonizes advanced machine learning techniques and practical cybersecurity solutions, setting a strong foundation for future developments in multi-modal network security frameworks. Integrating NL semantics enhances detection capabilities and offers a more intuitive understanding of network behaviors, crucial for cybersecurity professionals in diagnosing and responding to threats effectively. Future work includes analyzing \PacketCLIP's versatility by applying it to a broader range of network security tasks and exploring its performance in diverse network environments.








\bibliographystyle{IEEEtran}
\bibliography{IEEEabrv,all_bib}


\end{document}